\documentclass[pra,aps,twocolumn,showpacs,superscriptaddress]{revtex4}
\usepackage{graphicx}
\usepackage{multirow}
\usepackage{amsmath}
\usepackage{amssymb}
\usepackage{amsthm}
\usepackage{eucal}
\usepackage{bm}
\usepackage{upgreek}
\usepackage{framed}
\usepackage{mathrsfs}
\usepackage{latexsym}
\usepackage{float}
\usepackage{subfigure}

\newcommand{\ket} [1] {\vert #1 \rangle}

\newcommand{\hr}[1]{\hat{\rho}_{#1}}
\newcommand{\bs}[1]{\boldsymbol{#1}}

\begin{document}
\title{Quantum-optical channels that output only classical states}
\author{Krishna Kumar Sabapathy}
\email{krishnakumar.sabapathy@gmail.com}
\affiliation{F\'{i}sica Te\`{o}rica: Informaci\'{o} i Fen\`{o}mens Qu\`{a}ntics, Universitat Aut\`{o}noma de Barcelona, ES-08193 Bellaterra (Barcelona), Spain.}

\begin{abstract}
The Glauber-Sudarshan diagonal `weight' function provides a natural divide between  the quantum-optical notion of classical and nonclassical states of continuous variables systems.  Based on this demarcation, a channel is said to be nonclassicality breaking if it outputs only  classical states for any input state.  We focus on  multimode  bosonic Gaussian channels and classify those that are nonclassicality breaking by introducing a  criterion that needs to be satisfied by the matrices representing these channels.  The criterion can be interpreted as a nonclassicality benchmark for these channels since it quantifies the threshold noise at which there is a complete nonclassical to classical transition of the output states, i.e., it quantifies the robustness of the nonclassicality of the outputs of the channel against Gaussian noise. We then prove a striking `duality' between nonclassicality breaking and entanglement breaking bosonic Gaussian channels.
\end{abstract}
\maketitle
\section{Introduction}
The Gaussian domain of continuous variable systems, that includes Gaussian states and channels, plays a crucial role due to its experimental viability for quantum information and communication protocols especially in the field of  quantum optics\,\cite{rev1,rev2}.  Bosonic Gaussian channels (BGCs) are a special class of channels that are relevant for describing many physical systems like light transmission in optical fibers\,\cite{rev2}, certain quantum memories\,\cite{mem}, and some phenomena in gravitation\,\cite{adami}. Further, the communication capacities and related entropic properties of bosonic Gaussian channels have been of great current interest\,\cite{entropic,entropic2,gppt}. 

As with any quantum process, noise effects are ubiquitous\,\cite{books}, and of special interest to us is the role played by Gaussian noise\,\cite{hall94}. Various aspects of its manifestation in quantum protocols have been studied\,\cite{noiseeffects}. Much attention has been directed to the study of degradation of entanglement and nonclassicality in noisy environments\,\cite{degradation,ncb}. We focus our attention on the role played by Gaussian noise on the nonclassical characteristics of output states of a bosonic Gaussian channel. We quantify the threshold noise at which there is a complete nonclassical to classical transition of the output states, or equivalently, when the channel is rendered nonclassicality breaking. This turns out to be very useful since nonclassicality is  a key resource in quantum protocols, generation of entanglement\,\cite{resource} and superactivation of quantum communication\,\cite{gppt,wolf13} being important examples.  
   
The quantum-optical notion of classical states is well established based on the Glauber-Sudarshan diagonal weight function for continuous variables systems\,\cite{ecg,glauber}. Inspired by the definition of classical states, the notion of nonclassicality breaking channels was introduced in Ref.\,\cite{ncb}, though some examples were observed earlier in Refs.\,\cite{kraus10,agarwal}. Necessary and sufficient conditions for  single-mode bosonic Gaussian channels to be nonclassicality breaking were derived in Ref.\,\cite{ncb}. In this article we resolve the question of when a multimode bosonic Gaussian channel is nonclassicality breaking by obtaining a condition the associated noise matrix of a bosonic Gaussian channel has to satisfy and also present the various important implications of this result.

The outline for the rest of the article is as follows\,: in Section II we briefly introduce general phase space methods that are used in the later sections, in Section III we recall some useful notions of Gaussian states and channels which is our main focus, in Section IV we derive our first main result of the characterization of all multimode bosonic Gaussian channels that are nonclassicality breaking in terms of a criterion to be satisfied by the corresponding $(X,Y)$ matrices (Theorem 1) analogous to criteria previously known for Gaussian channels that are positive under partial transpose (PPT) [we also interchangeably choose to call these channels as NPT breaking (NPTB) channels] or entanglement breaking (EB), in Section V we show a close `dual' relationship between nonclassicality breaking and entanglement breaking bosonic Gaussian channels (Theorem 2), and finally, we conclude in Section VI.    

\section{phase space description}
A state $\hr{}$ of a quantum mechanical system  can be faithfully described 
by any  member of the one-parameter family of $s$-ordered quasi-probability distributions
or, equivalently, by the corresponding $s$-ordered characteristic functions\,\cite{cahill}. 
For $n$ modes of an electromagnetic field with quadrature operators
$(\hat{x}_i,\hat{p}_i),\,i=1,\cdots,n$ satisfying the 
commutation relation $[ \hat{x}_i, \hat{p}_j]=i\Omega_{ij}$, 
the $s$-ordered  characteristic function associated with a state $\hat{\rho}$
is defined as\,\cite{cahill}
\begin{eqnarray}
\chi_{s}(\boldsymbol{\xi}; \hat{\rho}) = \exp \left[\,\frac{s}{2} |\boldsymbol{\xi}|^{2}\,\right]\, {\rm 
Tr}\left[ \hat{\rho}\, \mathcal{D}({\boldsymbol{\xi}}) \right], \,\,\,\,\, -1 \leq s \leq 1.
\label{e1}
\end{eqnarray}
Here $\boldsymbol{\xi} = (\xi_1,\xi_2,\cdots,\xi_{2n})^T \,\in \, \mathbb{R}^{2n}$, $\mathcal{D}({\boldsymbol{\xi}}) =  \exp [-i\sqrt{2}\,\boldsymbol{\xi}^T \boldsymbol{R}]$ are the unitary multimode phase space Weyl-Heisenberg displacement operators, $\bs{R}= (\hat{x}_1,\hat{p}_1,\,\cdots,\,\hat{x}_n,\,\hat{p}_n)^T $,  $\Omega = \oplus_{i=1}^n i\sigma_2$ is the $n$-mode symplectic metric where $\sigma_2$ is the antisymmetric Pauli matrix, and $s \in [-1,1]$ is the order parameter. The special cases of $s= 1, 0, -1$ correspond to the normal ordering `$N$', symmetric ordering `$W$', and antinormal ordering `$A$' of the mode operators, respectively. Further, it immediately follows from\,(\ref{e1}) that the characteristic functions of a state $\hat{\rho}$ for two different values $s_1$, $s_2$ of $s$ are related as  
\begin{eqnarray}
\chi_{s_1}(\bs{\xi};\hr{})=
{\rm exp}\left[-\frac{(s_2-s_1)\,|\bs{\xi}|^2}{2}\right] \chi_{s_2}(\bs{\xi}; \hr{}).
\label{e2}
\end{eqnarray}

By Fourier transforming  the $s$-ordered characteristic 
function $\chi_{s}(\bs{\xi}; \hr{})$, we obtain
\begin{eqnarray}
\!\!\!W_{s}(\bs{\alpha};\hr{})\! =  
\frac{1}{(2\pi)^n}\int d^{2n} \bs{\xi}~{\rm exp}
[i\sqrt{2} \bs{\alpha}^T \bs{\xi}\,]~
\chi_{s}(\bs{\xi}; \hr{}).
\label{e3}
\end{eqnarray}
The quasiprobabilities $W_{s}$ corresponding to $s=-1,0,1$ are commonly known as the $Q$ function ($W_{-1}$), the Wigner function ($W_0$), and the diagonal `weight' function ($W_{1}$) (also called the Glauber-Sudarshan $P$ or $\phi$ function), respectively. The characteristic functions corresponding to $Q$, Wigner, and $\phi$ functions will be denoted by $\chi_{A},\,\chi_{W},\,\chi_N$, respectively. For the rest of the article we simply use only the `A', `W', `N' subscripts for the respective characteristic functions. The $Q$ function $Q(\bs{\alpha};\hr{})=\langle \bs{\alpha} | \hat{\rho} | \bs{\alpha} \rangle$, which 
by definition is manifestly pointwise nonnegative over the phase space ${\mathbb{R}^{2n}} \eqsim \mathbb{C}^n$, 
is a genuine probability distribution, a crucial fact which we exploit later. 

Any density operator $\hat{\rho}$ representing some state 
of $n$ modes of 
radiation field can always be expanded as\,\cite{ecg}
\begin{eqnarray}
\hr{} = \int\frac{d^{2n} \bs{\alpha}}{\pi^n}\, {\phi} (\bs{\alpha};\hr{}) | \bs{\alpha} \rangle \langle \bs{\alpha} |,
\label{e4}
\end{eqnarray}
where ${\phi} (\bs{\alpha};\hr{})= W_{1}(\bs{\alpha};\hr{})$ is the diagonal `weight' function
and $\{| \bs{\alpha} \rangle\}$ being the over-complete set of 
coherent states. 
A state $\hr{}$ is said to be classical if it can be expressed as a convex mixture of coherent states, i.e., 
\begin{eqnarray}
\hr{}\,\,{\rm is}\,\,{\rm classical}\,\Longleftrightarrow\, {\phi} (\bs{\alpha};\hr{}) \geq 0 \,\,\,\,{\rm for}\,\,\,{\rm all} \,\,\,\bs{\alpha}
\in \mathbb{R}^{2n}.
\label{e5}
\end{eqnarray}      
We now recall some useful properties of Gaussian states and channels.

\section{Gaussian states and channels}
A state $\hat{\rho}$ is said to be Gaussian if  its $s$-ordered characteristic function is Gaussian.
The symmetric or Weyl-ordered characteristic 
function\,($s=0$) corresponding to a Gaussian state has the form\,\cite{simon87,krp} 
\begin{eqnarray}
\chi_{W}(\bs{\xi};\hr{}) = \exp\left[-\frac{\bs{\xi}^T V \bs{\xi}}{2} \right],
\label{e6}
\end{eqnarray} 
where $V$ is the covariance matrix of the state $\hr{}$ (assumed to have vanishing first moments) and is defined as $V_{ij} = \langle \{\bs{R}_i,\,\bs{R}_j \} \rangle$. $V$ is real, $V=V^T,\,V>0$, and necessarily obeys the multimode uncertainty relation\,\cite{simon94} 
\begin{eqnarray}
V + i \Omega \geq 0. 
\label{e7}
\end{eqnarray}
Note that the chosen convention is such that the covariance matrix of the vacuum state is $1\!\!1_{2n}$ (we shall drop the subscript $2n$).
A necessary and sufficient condition for a Gaussian state to be classical is given by\,\cite{simon94}
\begin{align}
V\geq 1\!\!1.
\label{e7b}
\end{align}

A bosonic Gaussian channel (BGC) is a channel that maps every Gaussian state to a Gaussian state.  
Under the action of a bosonic Gaussian channel described by real matrices $(X, Y)$, $Y=Y^T,\,Y\geq 0$, the covariance matrix $V_{\rm in}$ corresponding to an input Gaussian state transforms as\,\cite{werner01}
\begin{eqnarray}
V_{\rm in} \rightarrow V_{\rm out} = X^T V_{\rm in} X +Y.
\label{e8}
\end{eqnarray} 
For an arbitrary input state $\hr{{\rm in}}$ the action of a bosonic Gaussian channel represented by $(X,Y)$ at the level of the symmetric-ordered characteristic function $\chi_{W}$ is given by 
\begin{eqnarray}
 \chi_{W}(\bs{\xi};\hr{{\rm out}}) = 
\chi_{W}(X \bs{\xi};\hr{{\rm in}}) \exp\left[-\frac{\bs{\xi}^T
    Y \bs{\xi}}{2} \right],
\label{e9}
\end{eqnarray}
and $(X,Y)$ has to satisfy the CP condition\,\cite{GCP} (we always assume trace-preserving condition in this article)
\begin{align}
Y + i\Omega \geq iX^T \Omega X.
\label{e10}
\end{align}
A bosonic Gaussian channel is known to be NPT breaking (NPTB) [recall that these channels are commonly referred to as PPT channels] if and only if  its corresponding $(X,Y)$ satisfies the condition\,\cite{gppt} 
\begin{align}
Y -i\Omega \geq iX^T \Omega X,
\label{e11}
\end{align}
and entanglement breaking (EB)\,\cite{holevo08}  if and only the noise matrix $Y$ can be decomposed into $Y=Y_1 + Y_2$ such that
\begin{align}
Y_1  + i\Omega \geq 0, ~~\text{and}~~ Y_2 \geq i X^T \Omega X.
\label{e12}
\end{align}

\section{Criterion for multimode nonclassicality breaking bosonic Gaussian channels}
We first begin with the notion of nonclassicality breaking channels. 

\noindent
{\em Definition}\,\cite{ncb}\,:  A channel $\Lambda$ is said to be nonclassicality breaking (NB) if and only if the output $\hr{}^{\,'} = \Lambda_{}(\hr{})$ is classical for every input state $\hr{}$. 

We wish to emphasize that this is a single-party notion unlike the NPT breaking and entanglement breaking case. Further, by definition, the set of nonclassicality breaking channels is convex.

We now give a characterization of bosonic Gaussian channels that are nonclassicality breaking in terms of the corresponding $(X,Y)$ matrices by using phase space methods mentioned in the earlier Sections. We wish to recall that the characteristic functions corresponding to the $Q$, Wigner, and $\phi$ functions are denoted by $\chi_A$, $\chi_W$, and $\chi_N$, respectively. Before we proceed to the main theorem we first prove a Lemma that will be used in the proof of the theorem.\\

\noindent
{\bf Lemma 1}\,:
Consider an additive classical noise channel, i.e., $X= 1\!\!1$ and $Y \geq 0$. Let $Y = 1\!\!1 + Y_0$, then $Y_0 + i\Omega \geq 0$ is a sufficient condition to render the channel nonclassicality breaking. In other words $Y_0$ is a valid covariance matrix. \\

\noindent
{\em Proof}\,: By Eq.\,\eqref{e9} we have
\begin{align}
\chi_N({\bs \xi};\hr{\rm out})&= \chi_N(\bs{ \xi};\hr{\rm in}) \, \exp\left[-\frac{ \bs{ \xi}^T\,Y\,\bs{ \xi}}{2} \right] \nonumber\\
&=\chi_W(\bs{ \xi};\hr{\rm in}) \, \exp\left[-\frac{ \bs{ \xi}^T\,Y_0\,\bs{ \xi}}{2} \right].
\label{e19}
\end{align}
We now apply a symplectic transformation $S \in {\rm Sp}(2n,\mathbb{R}): \bs{\xi}\to S\bs{\xi}$ such that $S^T Y_0 S $ is rendered diagonal, as guaranteed by Williamson's theorem with  diagonal entries $\geq 1$\,\cite{williamson,simon94,simon99}. Let us write $S^T Y_0 S = 1\!\!1 + \Delta$, with $\Delta \geq 0$. 
So Eq.\,\eqref{e19} now reads
\begin{align}
\chi_N(S{\bs \xi};\hr{\rm out}) = \chi_W(S\bs{ \xi};\hr{\rm in})\,\exp \left[-\frac{\bs{\xi}^T \bs{\xi}}{2} \right] \, \exp\left[-\frac{ \bs{ \xi}^T\,\Delta\,\bs{ \xi}}{2} \right].
\nonumber
\end{align}
Let $U_S$ be the unitary (metaplectic) operator that induces the symplectic transformation $S$, and we have
\begin{align}
\chi_N(S{\bs \xi};\hr{\rm out}) = ~~&\chi_W(\bs{\xi}; U_S\,\hr{\rm in}\,U_S^{\dagger}) \,\exp \left[-\frac{\bs{\xi}^T \bs{\xi}}{2} \right]\nonumber\\
&~~~\times \, \exp\left[-\frac{ \bs{ \xi}^T\,\Delta\,\bs{ \xi}}{2} \right].
\label{e21}
\end{align}
We denote $U_S\,\hr{\rm in}\,U_S^{\dagger}$ by $\hr{}^{\,'}$, and so by Eq.\,\eqref{e21} we have
\begin{align}
\chi_N(S{\bs \xi};\hr{\rm out}) &= \chi_W(\bs{\xi}; \hr{}^{\,'}) \,\exp \left[-\frac{\bs{\xi}^T \bs{\xi}}{2} \right] \, \exp\left[-\frac{ \bs{ \xi}^T\,\Delta\,\bs{ \xi}}{2} \right] \nonumber\\
&=\chi_A(\bs{\xi}; \hr{}^{\,'})  \, \exp\left[-\frac{ \bs{ \xi}^T\,\Delta\,\bs{ \xi}}{2} \right].
\label{e22}
\end{align}
Now we apply the Fourier transform\,[Eq.\,\eqref{e3}] to Eq.\,\eqref{e22}, and we have that $\phi(S^T\bs{\alpha};\hr{\rm out})$ is the convolution of  $ Q(\bs{\alpha};\hr{}^{\,'})$ with a Gaussian of the correct signature. We see that the diagonal weight function of the output state evaluated at the point $S^T\bs{\alpha}$ is always nonnegative since the $Q$ function is always nonnegative.\,$\square$ \\

We now present our main result on the characterization of all multimode  bosonic Gaussian channels that are nonclassicality breaking. \\

\noindent
{\bf Theorem 1}\,:
A Gaussian channel described by real $2n \times 2n$ matrices $(X,Y)$ with $Y=Y^T,\,Y \geq 0$ is nonclassicality breaking if and only if the pair $(X,Y)$ satisfies the following inequality\,:
\begin{equation}
Y - 1\!\!1 \geq i X^T \Omega X. 
\label{e13}
\end{equation} 
{\em Necessary}\,: To obtain the necessary condition we choose suitable input states and ask for the diagonal `weight' function of the corresponding outputs  under the channel action to be everywhere nonnegative on the phase space. We first consider the action of the channel at the level of the characteristic function. We also first consider the case of non-singular $X$. By Eq.\,\eqref{e9} we have
\begin{align}
\chi_N({\bs \xi};\hr{\rm out})&= \chi_W(X\bs{ \xi};\hr{\rm in}) \, \exp\left[-\frac{ \bs{ \xi}^T\,(Y-1\!\!1)\,\bs{ \xi}}{2} \right].
\label{e14}
\end{align}
Since $X$ is non-singular we can rewrite Eq.\,\eqref{e14} as 
\begin{align}
&\chi_N(X^{-1}{\bs \xi};\hr{\rm out})  \nonumber \\
& = \chi_W(\bs{ \xi};\hr{\rm in}) \, \exp\left[-\frac{ \bs{ \xi}^T\,X^{-T}(Y-1\!\!1)X^{-1}\,\bs{ \xi}}{2} \right].
\label{e15a}
\end{align}
There is an allowed symmetry which is the  application of arbitrary canonical transformation on the input states under which all the conditions must still hold true. Let $U_S$ be the unitary (metaplectic) operator that induces the symplectic transformation $S \in {\rm Sp}(2n,\mathbb{R})$ on phase space variables. Applying this canonical unitary on the input state, by Eq.\,\eqref{e15a} we have
\begin{align*}
\chi_N(X^{-1}{\bs \xi};\hr{\rm out}) &= \chi_W(\bs{ \xi}; U[S]\,\hr{\rm in}\,U[S]^{\dagger})\nonumber\\
&~~\times \,\exp \left[-\frac{\bs{\xi}^T \,X^{-T}(Y-1\!\!1)X^{-1}\,\bs{\xi}}{2} \right]. 
%\label{e15b}
\end{align*}
But this just induces the transformation $\bs{\xi}\to S\bs{\xi}$ and we have 
\begin{align*}
\chi_N(X^{-1}{\bs \xi};\hr{\rm out}) &= \chi_W(S\bs{\xi};\hr{\rm in})\nonumber\\ 
& ~~\times \,\exp \left[-\frac{\bs{\xi}^T \,X^{-T}(Y-1\!\!1)X^{-1}\,\bs{\xi}}{2} \right].
%\label{e15}
\end{align*}
We now denote $SX^{-1}{\bs \xi}$ by ${\bs \xi}^{\,\prime}$ and we have   
\begin{align} 
\chi_N({\bs \xi}^{\,\prime};\hr{\rm out}) &= \chi_W(\bs{ \xi};\hr{\rm in})\,\exp \left[-\frac{\bs{\xi}^T Y^{\,\prime}\bs{\xi}}{2} \right],\nonumber\\
 Y^{\, \prime} &= S^{-T} \,X^{-T}(Y-1\!\!1)X^{-1}\, S^{-1}.
\label{e15}
\end{align}
This particularly simple form of Eq.\,\eqref{e15} turns out to be useful to interpret for our purposes. The other symmetry of application of an arbitrary passive transformation after the channel action can be absorbed into $\bs{\xi}^{\,\prime}$ in the LHS of Eq.\,\eqref{e15}.

We now obtain necessary conditions on $Y^{\,\prime}$ by choosing suitable $\hr{\rm in}$. We require that after Fourier transforming  the RHS of Eq.\,\eqref{e15}, it is necessarily nonnegative for all $\xi$. Let us choose $\hr{\rm in}$ in the set of Gaussian states and  let $V_{\rm in}$ be the corresponding covariance matrix associated with an input Gaussian state. We have a necessary condition that $V_{\rm in} + Y^{\,\prime} \geq 0$ for the Fourier transform to be everywhere nonnegative.  Let $|\nu\rangle$ be an eigenvector of $Y^{\,\prime} $. We can always choose a $V_{\rm in}$ such that $\langle \nu | V_{\rm in}| \nu \rangle \to 0$ without violating the generalised uncertainty principle of Eq.\,\eqref{e7}. Similarly, we can choose another input Gaussian state $V_{\rm in}^{\,\prime}$ such that $\langle \nu^{\,\prime} | V_{\rm in}^{\,\prime}| \nu^{\,\prime} \rangle \to 0$ for another eigenvector $\ket{\nu^{\,\prime}}$ of $Y^{\,\prime}$, and repeat this process for each of the eigenvectors of $Y^{\,\prime}$. So we have that $Y^{\,\prime} \geq 0$.   

Now we take $\hr{\rm in}$ to be a product of Fock states, i.e., $\hr{\rm in} = |n\rangle \langle n| = |m_1\rangle \langle m_1| \otimes |m_2\rangle \langle m_2| \otimes \cdots |m_n\rangle \langle m_n|$, $m_{i}$'s being arbitrary. From\,\cite{soto},\,\cite{lee} and\,\cite{lut}, we have  that if $Y^{\,\prime} = \alpha 1\!\!1$, then $\alpha$ is necessarily $\geq 1$.  So we have that $Y^{\,\prime}$ has at least one eigenvector corresponding to eigenvalue greater than $1$ for the specially chosen input\,\cite{simon94}. We can now suitably choose another $\hr{\rm in}^{\,\prime}$ (related to $|n\rangle \langle n|$) such that the second eigenvector necessarily corresponds to an eigenvalue greater than $1$. This can be achieved by a passive unitary transformation which changes $\hr{\rm in}$ but our $Y^{\,\prime}$ is fixed since the condition to be derived must hold for any input state we choose. We can now iteratively repeat this procedure such that all eigenvectors of $Y^{\,\prime}$ necessarily correspond to eigenvalues $\geq 1$, or in other words, $Y^{\,\prime}$ is necessarily $\geq 1$.  So by Eq.\,\eqref{e15} we have that 
\begin{align}
S^T X^{-T}(Y-1\!\!1)X^{-1} S &\geq 1\!\!1\nonumber\\
\Rightarrow X^{-T}(Y-1\!\!1)X^{-1}  &\geq S^{-T} S^{-1}.
\label{e16}
\end{align}
Since the symplectic transformation was arbitrary we have that\,\cite{simon94} 
\begin{align}
X^{-T}(Y-1\!\!1)X^{-1} \geq  i\Omega \nonumber\\
\Rightarrow Y - 1\!\!1 \geq  i X^T \Omega X.
\label{e17}
\end{align}
For the case of singular $X$ we define a new `perturbed' matrix  $X_{\epsilon} = X + \epsilon 1\!\!!$. For this non-singular $X_{\epsilon}$ we use the condition in Eq.\,\eqref{e17}, and finally expand $X_{\epsilon}$ and take ${\rm lim}\, \epsilon \to 0$ to obtain the necessary condition. We see that we recover the same condition as in Eq.\,\eqref{e17} and therefore it is necessary for all $X$. We wish to emphasize that we have used a collection of multimode Fock states (over and above the Gaussian states) as inputs for which the corresponding outputs have to be rendered classical to derive our necessary condition. We now move on to the sufficiency part of the proof of Theorem 1. \\

\begin{table*}
\begin{tabular}{|c|c|c|c|c|}
\hline
&{\bf Property of a} & {\bf Condition on}\,\,$V$  &  {\bf Analogous property of } & {\bf Condition on}\,\,$\mathcal{V}(X,Y)$  \\
&{\bf Gaussian state} &   &  {\bf a BGC } &  \\
\hline
1.& Uncertainty relation & $V  +  i \Omega  \geq 0$ & CP & $\mathcal{V}(X,Y) +  i \Omega  \geq 0$ \\
\hline
2.&Pure & $V -\Delta \in \bs{\Gamma_n}, \, \Delta \geq 0$ & Quantum-limited& $\mathcal{V}(X,Y-\Delta) \in \bs{\mathcal{G}_n},\,\Delta \geq 0$ \\
&(Extremals in state space) &$\Rightarrow \Delta =0$&  (Extremals in convhull\{CP\})& $\Rightarrow \Delta =0$\\
\hline
3.&Mixed &$V - \Delta \in \bs{\Gamma_n}, \, \Delta \geq 0$&Noisy& $\mathcal{V}(X,Y-\Delta) \in \bs{\mathcal{G}_n},  \Delta \geq0$\\
\hline
4.&PPT& $V_{\rm AB} + i [\Omega_A \oplus \pm\, \Omega_B] \geq 0$&NPTB& $\mathcal{V}(X,Y) \pm i\Omega \geq 0$\\
\hline
5.&Separable&$V_{AB} - (V_A \oplus V_B)\geq 0,\,V_{A(B)} \in \bs{\Gamma_n}$&EB& $\mathcal{V}(X,Y-Y_1)   \geq 0, \, Y_1 \in \bs{\Gamma_n}$\\
\hline
6.&Classical&$V \geq 1\!\!1$& NB& ${\mathcal{V}(X,Y) \geq 1\!\!1}$\\
\hline
\end{tabular}
\caption{Showing a comparison of the various fundamental notions for Gaussian states and Gaussian channels. Here $\mathcal{V}(X,Y) = Y - i X^T \Omega X  $ is the characteristic matrix associated with $(X,Y)$, $\bs{\Gamma_n} = \{V\,|\, V+ i\Omega \geq 0,\, V\,{\rm real},\,V=V^T\} $  is the set of all valid covariance matrices in $n$-modes, $\bs{\mathcal{G}_n} = \{\mathcal{V}(X,Y)\,|\, \mathcal{V}(X,Y) + i \Omega \geq 0,\,Y = Y^T,\,Y \geq 0,\, X,Y\, {\rm real} \}$  is the set of all valid characteristic matrices on $n$-modes, and convhull\{CP\} is the set of all channels on $n$-modes. Property 5 for Gaussian states was shown in Ref.\,\cite{werner-wolf} and extremality of quantum-limited bosonic Gaussian channels in Refs.\,\cite{kraus10, holevoext}. For each property of the state or channel the associated condition should be satisfied over and above the condition in property 1.}
\label{table2}
\end{table*}

\noindent
{\em Sufficient}\,:
We begin with the action of the bosonic Gaussian channel on $\chi_{W}(\bs{\xi};\hr{\rm in})$ as given in Eq.\,\eqref{e9}.
Let $Y$ satisfy Eq.\,\eqref{e13}, and we write $Y = 1\!\!1 + Y_0$ where $Y_0 \geq i X^T\,\Omega\,X$.
The proof of the sufficiency of Theorem 1 for the case of additive classical noise channels $(X = 1\!\!1, Y \geq 0)$ was presented in Lemma 1. 

Next we consider the case of non-singular $X$. Equation\,\eqref{e9} can now be rewritten as 
\begin{align}
\chi_N({\bs \xi};\hr{\rm out}) = \chi_W(X\bs{ \xi};\hr{\rm in}) \, \exp\left[-\frac{ \bs{ \xi}^T\, Y_0\,\bs{ \xi}}{2} \right].
\label{e24}
\end{align}
By relabelling $X\bs{\xi}$ by $\bs{\xi}$, Eq.\,\eqref{e24} is now written as 
\begin{align}
\chi_N(X^{-1}{\bs \xi};\hr{\rm out}) = \chi_W(\bs{\xi};\hr{\rm in}) \, \exp\left[-\frac{ \bs{ \xi}^T\,(X^{-T}\,Y_0\,X^{-1})\,\bs{\xi}}{2} \right].
\nonumber
\end{align}
Comparing the above equation and Eq.\,\eqref{e19}, and by Lemma 1, we have that the channel is nonclassicality breaking if $X^{-T} Y_0 X^{-1}$ is a valid covariance matrix. In other words $X^{-T} Y_0 X^{-1} + i \Omega \geq 0$ or $Y_0 \geq i X^T \Omega X$, where we have applied the complex conjugation to the inequality. By adding  $1\!\!1$ to $Y_0$, we recover the sufficiency of the criterion in Theorem 1. 

The final case left is that of singular $X$. Let us write $X_{\epsilon} = X + \epsilon 1\!\!1$ so that in the limit $\epsilon \to 0$ we recover the original $X$ matrix. We  apply the sufficiency of the criterion to the non-singular $X_{\epsilon}$, and we have that $Y-1\!\!1 \geq i X_{\epsilon}^T \Omega X_{\epsilon}$. Now we expand $X_{\epsilon}$ and take the limit $\epsilon \to 0$. We find that we recover the sufficiency of the condition in Theorem 1. Hence we see that sufficiency of the inequality in Theorem 1 is proved for arbitrary $X$.\,$\blacksquare$

\begin{figure*}
\centering
\scalebox{0.38}{
\includegraphics{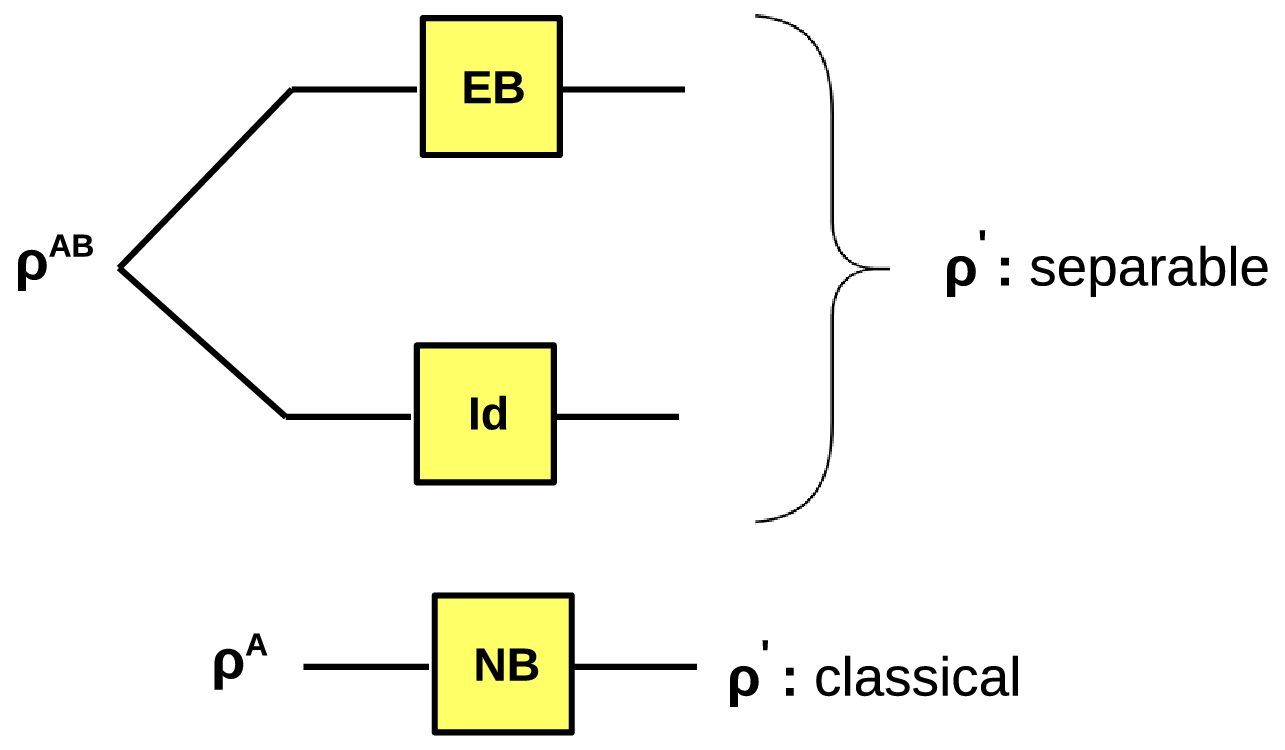}
\hspace{1cm}
\includegraphics{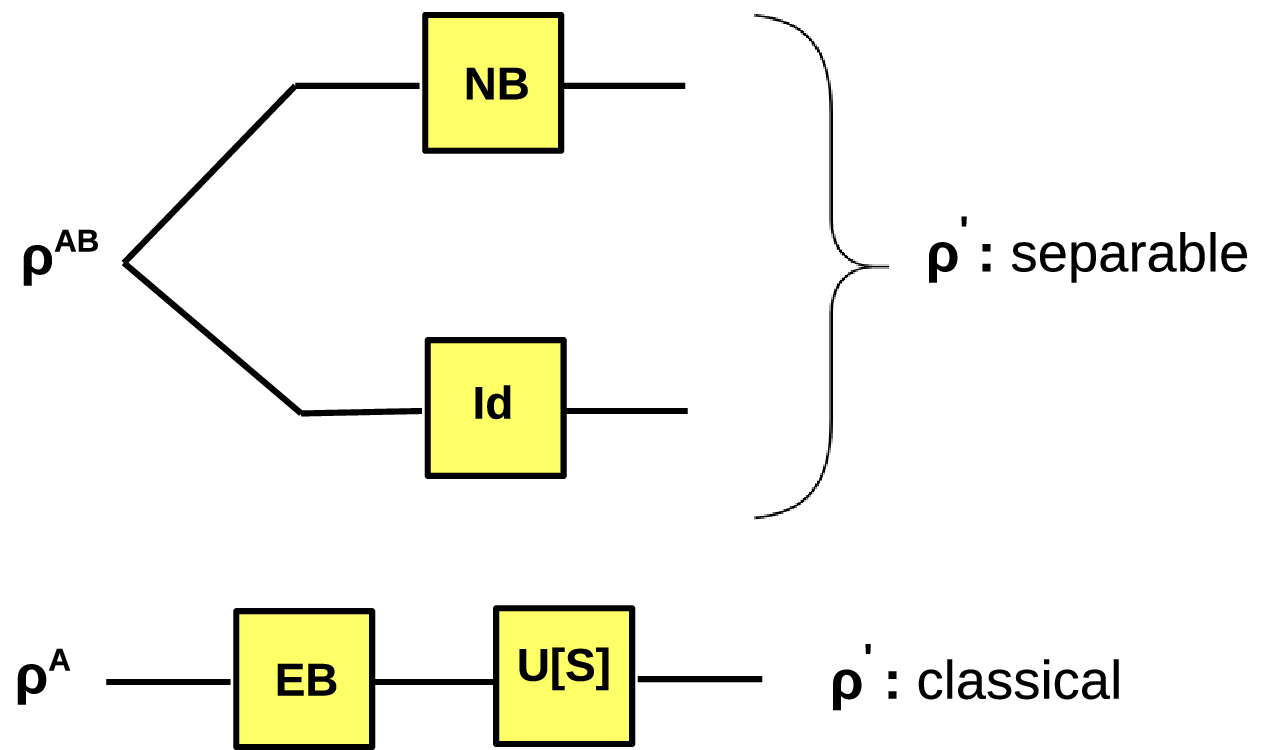}}
\caption{Depicting the `duality' between  entanglement breaking and nonclassicality breaking bosonic Gaussian channels. The figure on the left denotes the definitions of entanglement breaking and nonclassicality breaking channels. The diagram on the right brings out the following notion: Every nonclassicality breaking bosonic Gaussian channel is entanglement breaking whereas every entanglement breaking bosonic Gaussian channel  can be rendered nonclassicality breaking by following the channel action by a suitable active Gaussian unitary transformation $U[S],\, S \in {\rm Sp}(2n,\,\mathbb{R})$  which depends only on the channel parameters.  Here ${\rm Id}$ denotes the identity channel.\label{fig2}}
\end{figure*} 
\begin{table}
\begin{tabular}{|c|c|c|}
\hline
{\bf Type of}&{\bf Equivalence under}  &  {\bf Transformation}  \\
{\bf BGC}&{\bf pre/post-processing}&{\bf on (X,Y)}\\
\hline
CP,  NPTB,  EB &$U[S_1],\,U[S_2]$& $(S_1XS_2, S_2^TYS_2)$\\
\hline
NB&$U[S_1],\,U[R_2]$& $(S_1XR_2, R_2^TYR_2)$\\
\hline
\end{tabular}
\caption{Showing the equivalence properties of bosonic Gaussian channels (BGCs). Completely positive trace preserving (CP), nonpositive under partial transpose breaking (NPTB) and entanglement breaking (EB) are notions that are equivalent under pre- and post-processing by arbitrary Gaussian unitaries. Nonclassicality breaking (NB) channels are equivalent under pre-processing by arbitrary Gaussian unitaries whereas only passive transformations are allowed for post-processing. 
Here $S_1,S_2 \in {\rm Sp}\,(2n,\mathbb{R})$ and $R_1,R_2 \in \{{\rm Sp}\,(2n,\mathbb{R}) \bigcap {\rm SO}\,(2n)\}$.}
\label{table1}
\end{table}

\section{Connection between EB and NB bosonic Gaussian channels}
Having characterised all multimode bosonic Gaussian channels that are nonclassicality breaking, it is both instructive and transparent to introduce a new operator  $\mathcal{V}(X,Y)$  we call the {\em characteristic matrix} associated with $(X,Y)$ and define it  as 
\begin{align}
\label{e25b}
\mathcal{V}(X,Y) := Y - i X^T \Omega X.
\end{align} 
In terms of $\mathcal{V}(X,Y)$ the criterion of Theorem 1 is succinctly rewritten as 
\begin{equation}
\mathcal{V}(X,Y) \geq 1\!\!1, 
\label{e25c}
\end{equation}
analogous to the condition in Eq.\,\eqref{e7b} for a Gaussian state to be classical. We now rewrite the fundamental properties of bosonic Gaussian channels in terms of $\mathcal{V}(X,Y)$ and list them alongside the analogous notions for Gaussian states in Table.\,\ref{table2}. 
It is clearly seen that the criterion in Theorem 1 subsumes the NPT breaking condition and the connection to entanglement breaking channels is elucidated in Theorem 2. Further, the symmetry properties of our condition in Theorem 1 and indeed of other bosonic Gaussian channels is explained in Table.\,\ref{table1}. In passing we also note that NPT breaking is equivalent to entanglement breaking for single-mode bosonic Gaussian channels (by Simon's criterion\,\cite{simon00}) and multimode gauge-covariant bosonic Gaussian channels\,\cite{wolf13}.\\

\noindent 
{\bf Theorem 2}\,:
Every nonclassicality breaking  bosonic Gaussian channel is entanglement breaking. Every entanglement breaking bosonic Gaussian channel can be rendered nonclassicality breaking by composition with a suitable Gaussian unitary whose active component consists of only parallel single-mode canonical squeezing elements. In other words we have that every entanglement breaking bosonic Gaussian channel, say $\Phi^G_{\rm EB}$ ($G$ denoting Gaussian), when composed with a suitable Gaussian unitary can be made nonclassicality breaking, i.e., 
\begin{align}
\widetilde{\Phi}^G_{\rm NB} = U_G[S] \circ \Phi^G_{\rm EB}. 
\label{e26}
\end{align}
\noindent
{\em Proof}\,:  By Theorem 1, the noise matrix $Y$ of every nonclassicality breaking bosonic Gaussian channel can be decomposed as $Y = 1\!\!1 + Y_0$ where $1\!\!1 +  i\Omega \geq 0$ and $Y_0 \geq iX^T\, \Omega\, X$. By Eq.\,\eqref{e12} this  $(X,Y)$ corresponds to an entanglement breaking bosonic Gaussian channel. Hence every  bosonic Gaussian channel that is nonclassicality breaking is automatically entanglement breaking. 

For the second part of the proof let us consider an entanglement breaking bosonic Gaussian channel with $(X,Y)$ such that $Y=Y_1+Y_2$, $Y_1 + i\Omega \geq 0$ (being a valid covariance matrix), $Y_2 \geq iX^T \Omega X$. We apply a symplectic transformation $S\in {\rm Sp}\,(2n,\mathbb{R})$ that diagonalises $Y_1$ as guaranteed by the Williamson's theorem\,\cite{williamson,simon94,simon99}  so that $(X,Y) \to (\widetilde{X},\widetilde{Y})= (XS,S^TYS)$, and $\widetilde{Y}_1=S^TY_1S,\,\widetilde{Y}_2=S^TY_2S$. This transformation does not change the entanglement breaking property of the original channel. 

Since $Y_1$ is a valid covariance matrix, its symplectic eigenvalues are $\geq 1$. So we now have $\widetilde{Y}_1 \geq 1\!\!1$ and $\widetilde{Y}_2 \geq i \widetilde{X}^T \Omega \widetilde{X}$, where the second inequality involving $Y_2$ is covariant under the symplectic transformation. By Theorem 1 the channel $(\widetilde{X},\,\widetilde{Y})$ is now nonclassicality breaking.  Further, every symplectic matrix $S$ can be Euler decomposed as $S = R_1 \, D(\bs{\nu})\,R_2$, where $R_1,\,R_2$ are symplectic rotations and  $D(\bs{\nu}) = {\rm diag}(\nu_1,\nu_1^{-1},\nu_2,\nu_2^{-1},\cdots,\nu_n,\nu_n^{-1})$ is a positive diagonal matrix\,\cite{pramana}, the decomposition being inherently non-unique. $D(\bs{\nu})$ represents the active component consisting of parallel single-mode canonical squeezing elements.\,$\blacksquare$

A few remarks are in order. The relation in Eq.\,\eqref{e26} can be easily seen for entanglement breaking bosonic Gaussian channels that are already nonclassicality breaking. For example, using identity for the canonical unitary suffices, i.e., $\Phi_{\rm NB}^G = 1\!\!1 \circ \Phi_{\rm NB}^G$. 

The relation is however non-trivial for the case when we have an entanglement breaking bosonic Gaussian channel that is not nonclassicality breaking in the RHS of Eq.\,\eqref{e26}. Then the entanglement breaking bosonic Gaussian channel can be followed by the action of  suitable canonical unitary  to give rise to a nonclassicality breaking channel and the corresponding canonical unitary solely depends on the channel parameters of the entanglement breaking bosonic Gaussian channel as shown in Theorem 2.
 
One could also view the relation in Eq.\,\eqref{e26} in an equivalent way\,: every entanglement breaking bosonic Gaussian channel can be decomposed into a nonclassicality breaking channel followed by a suitable Gaussian unitary. This can be seen from the following. We start with a suitable nonclassicality breaking channel satisfying $\mathcal{V}(X,Y) \geq 1\!\!1$. From Theorem 1, let the  noise matrix be resolved as  $Y = Y_1 + Y_2$ such that  $Y_1 + i\Omega = 1\!\!1 + \Delta + i\Omega \geq 0,\,\Delta \geq 0$ and $Y_2 \geq iX^T \Omega X$. Further, we know that every variance matrix $V$ can always be decomposed as $S^TS + \Delta^{\,\prime}$, $\Delta^{\,\prime} \geq 0$\,\cite{simon94}. Then by suitably choosing the initial $\Delta$ and $Y_2$ for the nonclassicality breaking channel and applying a suitable Gaussian unitary, we can reach all entanglement breaking channels. This apparent role reversal of entanglement breaking and nonclassicality breaking bosonic Gaussian channels is depicted in Fig.\,\ref{fig2}. This is to be compared with the single-mode case where this structure was explicitly demonstrated\,\cite{ncb}. We make two final remarks regarding  the implications of Theorem 2. \\

\noindent
{\em Filterable states}\,: 
Let us consider a set of states denoted by ${\cal S}=\{\hr{1},\,\hr{2},\cdots \}$. We say that ${\cal S}$ is {\em filterable} if there exists a Gaussian unitary operation $U[{\cal S}]$ such that the set ${\cal S}^{\,'} =\{ U[{\cal S}]\,\hr{1}\,U[{\cal S}]^{\dagger},\,U[{\cal S}]\,\hr{2}\,U[{\cal S}]^{\dagger},\dots\}$ obtained by applying $U[{\cal S}]$ to the elements of ${\cal S}$ consists of only classical states. It was shown in\,\cite{ncb} that for every single-mode entanglement breaking bosonic Gaussian channel, denoted by $\Phi$, ${\cal S}= \{\Phi[\hr{}]\,| \,\hr{} \in  \text{state space}\}$ is filterable. In Theorem 2 we have proved  that the collection of output states for all input states of a multimode entanglement breaking bosonic Gaussian channel is filterable and that the corresponding Gaussian unitary transformation is solely dependent on the $Y_1$ partition of $Y$ corresponding to the entanglement breaking channel. \\

\noindent
{\bf Corollary}\,: 
The classical capacity of every multimode nonclassicality breaking bosonic Gaussian channel is additive and its quantum capacity is zero. \\
\noindent
{\em Proof}\,: This is a straightforward consequence of Theorem 2. It is well known that for every entanglement breaking channel the classical capacity is additive and its quantum capacity is zero\,\cite{shor02,holevo08}. The statement of the Corollary follows from the fact that every nonclassicality breaking bosonic Gaussian channel is automatically entanglement breaking. Hence the classical capacity is additive and the quantum capacity is zero for every nonclassicality breaking bosonic Gaussian channel.\,$\blacksquare$

\section{Conclusions}
We have classified all  multimode bosonic Gaussian channels  that are nonclassicality breaking, these being the analog of multimode Gaussian states that are classical.  The conditions for a bosonic Gaussian channel to be NPT breaking or entanglement breaking were previously known. The criterion for bosonic Gaussian channels to be nonclassicality breaking was derived in this article and is listed in Table\,\ref{table2}, alongside other fundamental properties of Gaussian states and their channel counterpart in very close analogy.  

\begin{figure}
\begin{center}
\scalebox{0.55}{\includegraphics{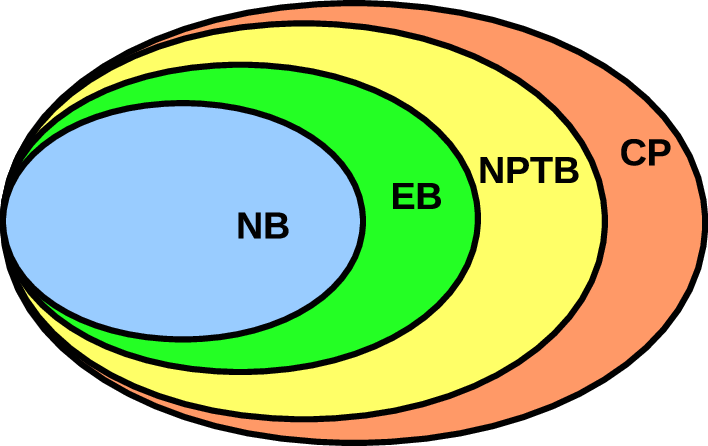}}
\end{center}
\caption{A schematic diagram depicting the (non-convex) sets of bosonic Gaussian channels (CP), and its properties like NPT breaking (NPTB), entanglement breaking (EB), and nonclassicality breaking (NB). The set of nonclassicality breaking channels and the set of extremal channels have a non-trivial intersection which includes, for example, single-mode quantum-limited phase conjugation channels (and taking $n$ copies thereof)\,\cite{ncb,kraus10,holevoext}.\label{fig3}}
\end{figure}

Further, we proved an interesting duality  that every nonclassicality breaking bosonic Gaussian channels is entanglement breaking and that every entanglement breaking bosonic Gaussian channel can be rendered nonclassicality breaking by the action of a Gaussian unitary whose active component consists only of parallel single-mode canonical squeezing elements; the unitary being dependent only on the channel parameters of the entanglement breaking  bosonic Gaussian channel. Therefore the set of nonclassicality breaking bosonic Gaussian channels is a  subset of the set of all entanglement breaking bosonic Gaussian channels. We depict the set-theoretic nature of the various notions of a bosonic Gaussian channel in Fig.\,\ref{fig3}. This special relationship between nonclassicality breaking and entanglement breaking bosonic Gaussian channels as shown in Theorem 2 is not known to exist for other pairs in the hierarchy depicted in Fig.\,\ref{fig3}.

We find that, in general, too much noise could be detrimental as it can render the channel nonclassicality breaking. In such a case the channel is not only entanglement breaking but it also only produces classical outputs and hence is ineffective for protocols aiming to exploit the power of nonclassical states. In effect, the condition in Theorem 1 can be interpreted as a kind of nonclassicality benchmark for bosonic Gaussian channels in the sense that every bosonic Gaussian channel satisfying this criterion is guaranteed to  produce only classical states at the output irrespective of the input. 

In other words, the nonclassical character of a bosonic Gaussian channel is quantified in terms of the noise it can tolerate beyond which all the output states are rendered classical. Since nonclassicality is a crucial resource for quantum protocols, such a characterization is of practical importance.  We believe that the results presented here have far-reaching implications for both theoretical and experimental aspects of realization  of quantum-optical networks\,\cite{kimble}, benchmarking\,\cite{bmarks}, continuous variable quantum key distribution, testing quantum sources, and other quantum-optical protocols\,\cite{rev1,rev2}. We refer the reader to\,\cite{nc} for an application of the tools and techniques developed in this article.

\section*{ACKNOWLEDGMENTS}
The author is very grateful to Andreas Winter, Rajiah Simon, J. Solomon Ivan, and R. Garc\'{i}a-Patr\'{o}n for many insightful discussions, Werner Vogel for useful feedback, and an anonymous AQIS'15 conference referee for suggesting the terminology NPT breaking in lieu of PPT channels. The author is supported by the ERC, Advanced Grant  ``IRQUAT'', Contract No. ERC-2010-AdG-267386, and Spanish MINECO  project FIS2013-40627-P, and the Generalitat de Catalunya CIRIT, project 2014-SGR-966.

\end{document}